\documentclass[twocolumn]{aastex61}
\pdfoutput=1 %for arXiv submission
\usepackage{amsmath,amstext,mathtools,graphicx,rotating}
\usepackage[T1]{fontenc}
\usepackage{apjfonts} 
\usepackage[figure,figure*]{hypcap}

\newcommand{\irissq}{$\mathbf{IRIS^{2}}\ $}
\newcommand{\noirp}{$\widetilde{RP}^{syn}_{t_{exp},noi}$}

\newcommand{\noirma}{$^{MC}\widetilde{RMA}^{syn}_{t_{exp},noi}$}

\newcommand{\mgii}{\ion{Mg}{2} h \& k}
\newcommand{\mguv}{\ion{Mg}{2} UV2\&3}

\newcommand{\synrp}{$RP^{syn}$}

\newcommand\nnoirma[1]{$^{[#1]}\widetilde{RMA}^{syn}_{t_{exp},noi}$}
\newcommand{\uncertp}{$\sigma_p$}
\newcommand{\chith}{$\chi^2_{threshold}$}

\shorttitle{Improved determination of uncertainty in IRIS$^2$}
\shortauthors{Sainz Dalda and De Pontieu}

\begin{document}

\title{Recovering Thermodynamics from Spectral Profiles observed by IRIS (II): improved calculation of the uncertainties based on Monte Carlo experiments}

\author[0000-0002-3234-3070]{Alberto Sainz Dalda}
\affil{Lockheed Martin Solar \& Astrophysics Laboratory, 3251 Hanover Street, Palo Alto, CA 94304, USA}
\affil{Bay Area Environmental Research Institute, NASA Research Park, Moffett Field, CA 94035, USA.}
\author[0000-0002-8370-952X]{Bart De Pontieu}
\affil{Lockheed Martin Solar \& Astrophysics Laboratory, 3251 Hanover Street, Palo Alto, CA 94304, USA}
\affil{Rosseland Center for Solar Physics, University of Oslo, P.O. Box 1029 Blindern, NO-0315 Oslo, Norway}
\affil{Institute of Theoretical Astrophysics, University of Oslo, P.O. Box 1029 Blindern, NO-0315 Oslo, Norway}

\correspondingauthor{Alberto Sainz Dalda}\email{sainzdalda@baeri.org\\asainz.solarphysics@gmail.com}

\begin{abstract}

Observations by the Interface Region Imaging Spectrograph (IRIS) in the \mgii\ spectral lines have provided a new diagnostic window towards the knowledge of the complex physical conditions in the solar chromosphere. Theoretical efforts focused on understanding the behavior of these lines have allowed us to obtain a better and more accurate vision of the chromosphere. These efforts include forward modeling, numerical simulations, and inversions. In this paper, we focus our attention on the uncertainties associated with the thermodynamic model atmosphere obtained after the inversion of the \mgii\ lines. We have used $\approx 50,000$ synthetic representative profiles of the \irissq\ database to characterize the most important source of uncertainties in the inversion process, {\it viz.}: the inherent noise of the observations, the random initialization of  process, and the selection criteria in a high-dimensional space. We have applied a Monte Carlo approach to this problem. Thus, for a given synthetic representative profile, we have created five randomized noise realizations (representative of the most popular exposure times in the IRIS observations), and inverted these profiles five times with different inversion initializations. The resulting 25 inverted profiles, fits to noisy data, and model atmospheres are then used to determine the uncertainty in the model atmosphere, based on the standard deviation and empirical selection criteria for the goodness of fit. With this approach, the new uncertainties of the models available in the \irissq\ database are more reliable at the optical depths where the \mgii\ lines are sensitive to changes in the thermodynamics.
\end{abstract}


\section{Introduction}

The study of the chromosphere is critical to understand the solar atmosphere \citep{Carlsson19}. Although this region of the solar atmosphere has been observed for decades, 
 understanding it is still a challenge. This is due to several major issues: 1. the complex coupling between the radiation field and the local magnetic field and thermodynamic conditions, which means interpretation of radiation must consider non-local thermodynamic equilibrium; 2. the transitions from fully ionized plasma to partially ionized and back to fully ionized, and from dominated by the plasma to domination by the magnetic field; 3. the highly dynamic and highly structured nature of the chromosphere on small spatio-temporal scales, necessitating sub-arcsecond high-quality observations on timescales of seconds. During the last few decades, both theoretical and observational improvements have allowed us to gain a better knowledge of the chromosphere and the events that occur in this region \citep[e.g.][]{Scharmer08, Vissers15b,Leenaarts11,Leenaarts13a, Leenaarts13b, DePontieu14a,QuinteroNoda16b, delaCruzRodriguez19,Carlsson19,Centeno21,DePontieu21,Ishikawa21,Vissers22,TrujilloBueno22}.
 
 The Interface Region Imaging Spectrograph (IRIS, \citealt{DePontieu14a}) has been providing high-resolution observations (free from seeing effects introduced by the Earth's atmosphere) of the chromosphere through the near-ultra violet spectral range around the \mgii\ lines since 2013. The IRIS wavelength range also contains the \ion{Mg}{2} UV triplet lines (hereafter denoted as \mguv). The \mgii\ lines are optically thick lines, being sensitive to the conditions at the high- and mid- chromosphere \citep{Leenaarts13a, Leenaarts13b, Pereira13}, while the \mguv\ lines typically form lower in the chromosphere (although under flaring conditions the line formation may be different \citealt{Kerr16,RubiodaCosta16}). The most reliable method to derive physical information along the optical depth from these lines is by the "inversion" of these lines. This involves an iterative process in which, at first, an initial atmosphere is assumed and the radiative transfer equations are solved considering the non-local thermodynamic equilibrium and the partial frequency redistribution of the radiation from scattered photons, leading to a refinement in the underlying atmosphere, followed by further iterations. 
 
 The state-of-the-art Stockholm Inversion Code (STiC, \citealt{delaCruzRodriguez16, delaCruzRodriguez19}) is the only available code capable of inverting these lines under these conditions. However, the inversion of a single \mgii\ profile is computationally expensive (2.5 CPU-hour per observed profile). To minimize this burden, we created the IRIS Inversion based on Representative profiles Inverted by STiC (\irissq, \citealt{SainzDalda19}). This technique is based on the inversion of the representative profiles (RP) of a broad selection of observations taken by IRIS in the \mgii\ lines. A RP is the averaged profile of those profiles belonging to a data set that share the same shape, i.e. a similar distribution of the intensity over a given spectral range. This shape - or profile - is the signature of the conditions in the solar atmosphere where the radiation comes from. Therefore, the RP is the average of those profiles sharing similar conditions in the Sun. It is natural to define a Representative Model Atmosphere for the atmospheric conditions associated with the RP, which is obtained from the inversion of the RP. The core of \irissq is the \irissq database, which has 3 components: i) the synthetic RPs (\synrp); ii) their corresponding RMAs obtained from the inversion of RPs; and iii) the uncertainty of the thermodynamics variable $p$,  \uncertp,  of the RMA associated with the observed \synrp. The \irissq database consists of $\approx$ 50,000 items, obtained from: 1. clustering 312 data sets on different targets (observed by IRIS) by using the $k-means$ technique \citep{Steinhaus57,MacQueen67}; 2. inverting each RP with STiC; 3. obtaining the \synrp, RMAs and $\sigma_{p}$. The physical information relies on the relationship between the \synrp, the RMA, and the uncertainties of the latter (\uncertp), while the statistical significance of \irissq is given by the selection of the datasets considered in the database, which takes into account different solar features, exposure times, and locations on the solar disk.

% \begin{table}
%\begin{center}
% \begin{tabular}{ccc}
% $\mu$ & No. RP & No. RP [\%] \\
% 	\hline
%  0.05 & 2069 & 4.1 \\
% 0.15 & 1744 & 3.4 \\
% 0.25 & 798 & 1.6 \\
% 0.35 & 1755 & 3.5 \\
% 0.45 & 1118 & 2.2 \\
% 0.55 & 3031 & 6.0 \\
% 0.65 & 2552 & 5.0 \\
% 0.75 & 4944 & 9.8 \\
% 0.85 & 7029 & 13.9 \\
% 0.95 & 25559 & 50.5 \\
% \hline
% \end{tabular}
%\caption{Distribution of Representative Profiles (RP) in the \irissq\ database for $\mu-0.05 \le \mu < \mu+0.05$.}
%\end{center}
%\end{table}

  \begin{table*}
 	\begin{center}
 		\begin{tabular}{lcccccccccc}
 			$\mu$ & 0.05 & 0.15 & 0.25 & 0.35 & 0.45 & 0.55 & 0.65 & 0.75 & 0.85 & 0.95 \\
 			\hline
 			No. RP & 2069 & 1744 & 798 & 1755 & 1118 & 3031 & 2552 & 4944 & 7029 & 25559 \\
 			No. RP [\%] & 4.1 & 3.4 & 1.6 & 3.5 & 2.2 & 6.0 & 5.0 & 9.8 & 13.9 & 50.5  \\
 		    \hline
 		\end{tabular}
 		\caption{Distribution of Representative Profiles (RP) in the \irissq\ database for $\mu-0.05 \le \mu < \mu+0.05$.\label{table:mu}}
 	\end{center}
 \end{table*}

 In the first publicly released version of \irissq, the uncertainty of a physical variable \uncertp\ was obtained using the expression\footnote{A formal derivation of this expression can be found by using the equations of section 2.3 in \citealt{SanchezAlmeida97b} and of sections 6.2 and 6.3 in \citealt{BellotRubio98c}.}  (see and  \citealt{DelToroIniesta16}):
 
 \begin{equation}
 \sigma^{2}_p = \frac{2}{nm+r}\frac{\sum_{i=1}^{q}{\left[ I(\lambda_i)^{obs} - I(\lambda_i; \mathbf{M})^{syn\  RP@STiC})^{2}\right] \frac{w_{i}^2}{\sigma_{i}^2}}}{\sum_{i=1}^{q}R^2_{p}(\lambda_{i})\frac{w_{i}^2}{\sigma_{i}^2}}\label{Eq:unc}
 \end{equation}

with $i = 0,..., q$ the sampled positions in the wavelength $\lambda_{i}$,
%spectral positions in the profile
$w_{i}$ their weights, $\sigma_{i}$ the uncertainties of the observation (e.g. photon noise), $m$ the number of physical quantities in the model $\mathbf{M}$ evaluated in $n$ grid points along the solar atmosphere, $r$ the number of physical quantities considered constant along that atmosphere, and $R_{p}$ the {\it Response Function} (RF) of a Stokes parameter to the physical quantity $p$ \citep{Mein71,LandiDegl'Innocenti79,RuizCobo92}. The RF provides the sensitivity of a wavelength sample in a Stokes profile to changes of a physical quantity. In this study, we only consider the Intensity Stokes parameter, $I$.

The expression above is valid to calculate \uncertp, however, practical cases using \irissq\ show an underestimation in \uncertp\ in those regions where the line is sensitive to the changes in physical variable $p$, and an overestimation of \uncertp\ where the line is not so sensitive to those changes. This is due to the fact that the $R_p$ is  calculated considering all the optical depths (or nodes) of the model $\mathbf{M}$, while for $I(\lambda_i; \mathbf{M})^{syn\  RP@STiC}$ only variations in a selected number of nodes in the model $\mathbf{M}$ are considered. That means, $R_p$ (RF) encodes the information in all the optical depths, while the \synrp  comes from a model evaluated in selected optical depths.  Thus, in the particular nodes where the line is more sensitive to changes in $p$, the $R_p$ will be larger than in those nodes where it is less sensitive, making the \uncertp\ ($\sim R_p^{-1}$) smaller in regions where the line is more sensitive and larger in the regions where the line is less sensitive. This is the expected behavior, but in practice, in many cases, the obtained \uncertp\ is very low (high) for the optical depths where the line is (not) sensitive to changes in the physical variable $p$.
 
In this paper, we present a new approach to calculate the uncertainties of the RMAs in the \irissq database initially presented by \cite{SainzDalda19}. In Section \ref{sec:methodology}, we explain how these new uncertainties have been calculated using a Monte Carlo simulation approach. The criteria used to determine the uncertainties are presented in Section \ref{sec:selcriteria}. In Section \ref{sec:discussion}, we evaluate the results obtained with the new version of \irissq with those obtained from inversion using STiC. Finally, in Section \ref{sec:conclusions} we present the main conclusions and limitations of the new \irissq database. %In the extension side, we present two new inversion techniques based: one based in the k-nearest neighbor method (Section \ref{sec:knn}); and other able to predict the thermodynamics and the synthetic profile associated to an observed \mgii profile by an invertible neural network (Section \ref{sec:inn}). 

\section{Methodology}\label{sec:methodology}

When we invert an observed profile there are several factors that introduce a randomness to some key elements in the inversion. First, the noise inherent to an observation, both the one associated to the distribution of photons {\it detected} by the instrument (i.e., Poisson noise for our NUV photons), and the one associated to the readout or other electronic variations in our detector. 

In addition, the initialization of the iterative inversion process is usually randomized. Thus, the {\it path} started and followed during an inversion of an observed profile may be different from another independent inversion for the same profile, which may yield different results. To better understand the impact of this randomness in the initialization of the inversion, we can invert the same profile several times with different initializations. This {\it Monte Carlo inversion} approach to quantify the uncertainty was used for the first time, to the best of our knowledge, by \citealt{WestendorpPlaza99}. Another source of possible variability in the inversion results comes from the initial atmosphere model. To minimize this, the inversion code DeSIRe \citep{RuizCobo22} uses several initial guess models to independently invert the same profile, selecting the best fit of all the fits prociced by each inversion. We have not considered this case in our study since the \irissq database was built with the results from the inversion of the RPs using an unique initial guess model (FALC, \citealt{Fontenla93}), and that is the one we only consider in our Monte Carlo approach.

One other aspect to consider when estimating uncertainties is that the inversion technique is based on the minimization of

\begin{equation}\label{eq:chi2}
\chi^{2} = \frac{1}{\nu}\sum_{i=0}^{q}{(I(\lambda_i)^{obs} - I(\lambda_i, \mathbf{M})^{syn\  RP@STiC})^{2}\frac{w_{i}^2}{\sigma_{i}^2}} 
\end{equation}

with $i = 0,..., q$ the sampled wavelengths,$w_{i}$ their weights, $\sigma_{i}$ the uncertainties of the observation (e.g. photon noise)\footnote{Formally, the Equation \ref{eq:chi2} consider a weight and a noise per spectral position per profile. However, for computational reasons only one weight and noise level per spectral profile is provided for all the profiles inverted in a batch. In this study, a batch is  all the \noirp\ at a given $t_{exp}$ at a given $\mu$.}, and $\nu$ the number of observables, i.e., the spectral samples. This is the weighted Euclidean distance between the observed (input) and the synthetic (output) profile, with the weight higher for those wavelengths that we are more interested in. However, as we will see below, this metric is not optimal for those cases where the dimension of the observation, i.e. the number of observed wavelengths, is high. 

The method that we have used to estimate the uncertainties associated with the \synrp-RMA takes into account all these issues.

\subsection{Building a Noisy Database}
As we have already mentioned the physical information in \irissq is given by the relationship between the \synrp\ and the RMA. This information is determined by the physical considerations made in solving the radiative transfer equation for the \mgii\ lines. Therefore, we can consider the \synrp - RMA pair as the {\it ground truth}. Keeping that in mind, we have created a new noisy database using these pairs as {\it guides}. The steps taken in this process are the following:

\begin{itemize}
	\item We applied Poisson noise to a \synrp at a given $t_{exp}$ (exposure time). The values of $t_{exp}$ are $1, 4, 8$ and $30 s$, which are the most used in the IRIS observations. We also add a readout noise characterized by a Gaussian distribution with a standard deviation of 18 ($e^-$) \citep{DePontieu14a}.
	The result is a noisy synthetic profile dependent on the exposure time, \noirp. 
	\item We repeat the previous step 5 times considering a different randomization for the same \synrp\ each time. Thus, we now get 20 noisy profiles associated with one of the \synrp\ in the \irissq database: 5 random realizations in noise for each of the 4 exposure times considered. We denote these profiles as \noirp, with the $\tilde{\ \ \ }$ indicating the noisy nature of the profile for a given $t_{exp}$, with the 5 randomizations in the noise $noi$. Thus,  $t_{exp} = [1,4,8,30] s$, and $noi = 1,...,5$. 
	\item Each \noirp\ is independently inverted 5 times with STiC, following the same inversion scheme as the one used in \irissq. This {\it Monte Carlo simulation} tries to characterize the impact of the randomness of the initialization of the inversion.  Hence, for each \noirp we obtained 5 \noirma. The superscript $MC$ indicates the 1,...,5  independent (initialization) inversions. The superscript $syn$ denotes that the associated input profile in the inversion is not an observed profile ($obs$), but a (noisy) synthetic profile. 
	\item At this point, for a given \synrp\ at a given exposure time, $t_{exp}$, we have 25 associated \noirma. Thus, each of the 25 \noirma\ takes into account the random nature of the noise ($noi$) for a given exposure time ($t_{exp}$), and the random nature of the initialization of the inversion ($MC$). 
\end{itemize}

\begin{figure*}[h!]
	\centering
	\includegraphics[width=.75\linewidth]{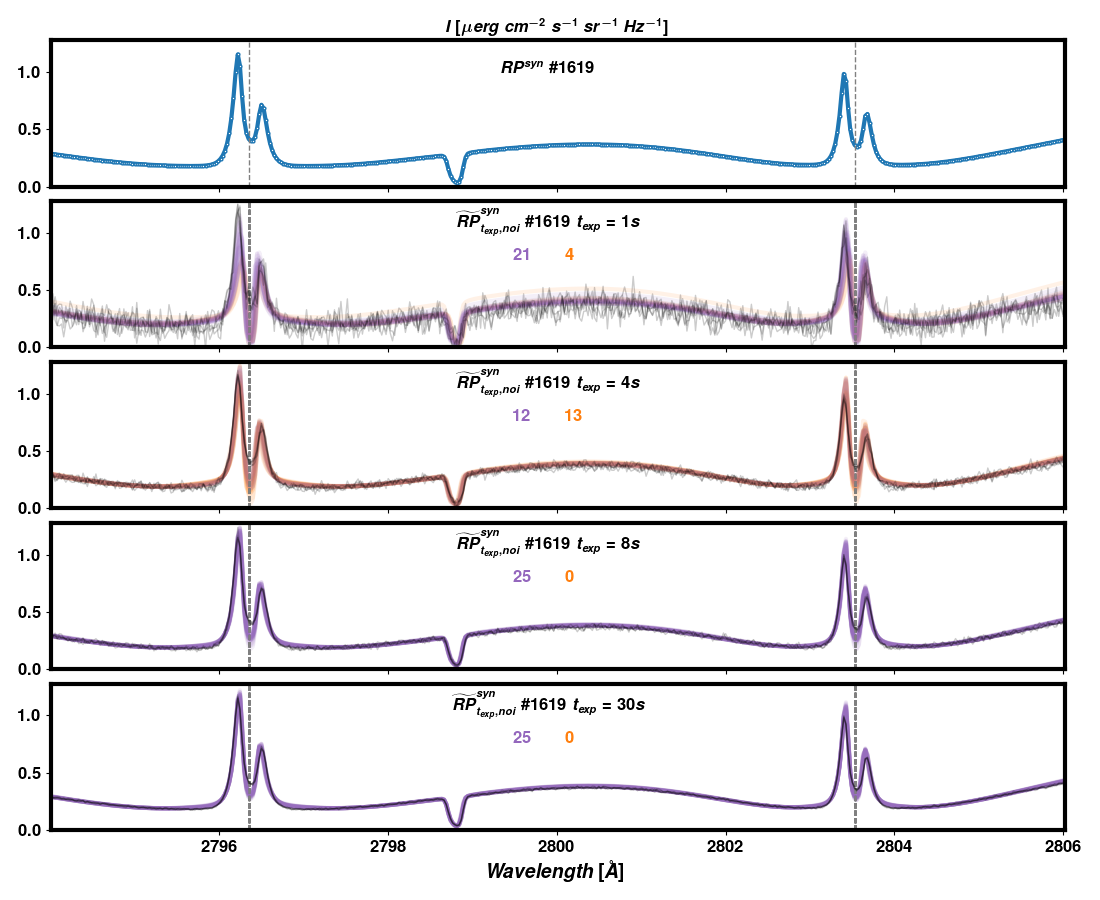}
	\includegraphics[width=.75\linewidth]{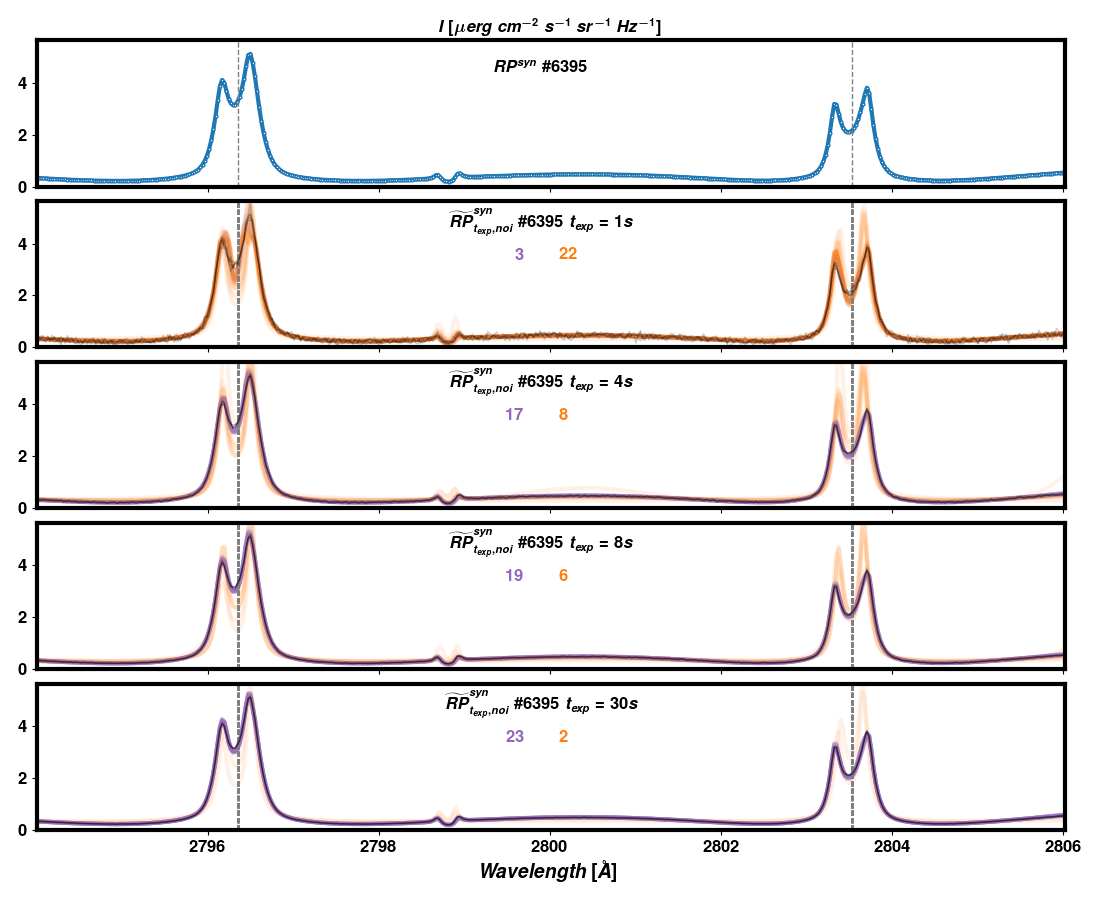}
	\caption{Top panel: The \synrp\ \#1619 is shown in the first row. The four following rows show their corresponding \noirp\ (in black) for $t_{exp}=1, 4, 8$ and $30s$, and the "good" and "bad" MC fits in violet and orange respectively (the colored numbers indicate the number of "good" and "bad" MC fits). Bottom panel: the same for \synrp\ \#6395.}
	\label{fig:invnoisy}
\end{figure*}

The new noisy database consists of $\approx$ 1.25$M$ (million) \noirp-\noirma\ pairs for each $t_{exp}$, or a total of 5$M$ pairs considering all the exposure times.  Figure \ref{fig:invnoisy} shows two examples of \noirp.  In both panels, the first row shows the \synrp\ as included in the \irissq database. The next 4 rows show in black  the 5 noisy profiles  for $t_{exp} = 1, 4, 8$ and $30s$, in violet the inverted synthetic profiles that fit the 5  \noirp\ with a $\chi^2 \le 3$ ("good" fits), and in orange those inverted synthetic profiles that fit the \noirp\ with a $\chi^2  > 3$ ("bad" fits). Because $noi = 1,...,5$, the total number of inverted profiles displays for a given \synrp\ at a given $t_{exp}$ is 25. The number of the {\it good}  and the {\it bad} inverted synthetic profiles  is given in each panel in violet and orange fonts. In the following section, we describe why we have selected this threshold for $\chi^2$. Each line is plotted with a transparency factor so that the intensity of the color expresses the probability of signals. Thus, the common values in each profile are more visible than those where the profiles are less common. This effect can be seen in the wavelength range between the two \mgii\ lines (which we refer to as the photospheric "bump", since it is formed in the photosphere) for the \noirp \#1619 (second panel from the top in Fig. \ref{fig:invnoisy}) for $t_{exp}=1s$, where the 4 "bad" inverted profiles (in orange ) show a contribution located at a range of different intensity values. As a result, the colored lines look rather faded in that spectral region. In contrast,  the "good" inverted profiles (in violet) overlap in this wavelength range, and also in the \mgii lines and in the \mguv\ lines. If we now look at the profiles for $t_{exp}=4s$, we can barely distinguish the "good" inverted profiles (12) from the bad ones (13), since they mostly contribute equally in the same spectral region with similar values, resulting in a brownish profile quite well defined in the photospheric bump and the \mguv\, but slightly blurred or dispersed in the \mgii\ lines. With this visualization we want to illustrate how various spectral regions contribute (or not) to the nature of the fit ("good" or "bad"), and thus to the uncertainty associated with the RMA. In Section \ref{sec:discussion} we discuss these inverted profiles,  but we have to first answer an important question: when do we consider a fit to be {\it good} or {\it bad}? %The answer to this question is given in detail in the following section. 
% TODO: \usepackage{graphicx} required

%For computational reasons, the \noirp from the data

\subsection{Selection Criteria}\label{sec:selcriteria}

The next step is to calculate the uncertainty associated with the \synrp-RMA pair. We use the Monte Carlo simulation to calculate the uncertainty for a physical variable as the standard deviation of all the Monte Carlo experiments,  that is, the standard deviation of the 25 \noirma\ associated with an \synrp. In this context, we refer to a Monte Carlo simulation as the exercise of calculating the 25 inversions for a given \synrp (5 times for each of the 5 \noirp\ associated with that \synrp) , and to a Monte Carlo experiment as one of these 25 inversions (or experiments). 

In an ideal scenario, we would need a large number of Monte Carlo experiments for each Monte Carlo simulation: this means a large number of independent inversions considering several random initializations of the noise for a given exposure time. In this fashion, the impact of statistical outliers would be reduced compared to our approach with just 25 simulations. However, such an approach is computationally very expensive and not practical. Our current approach to build the "noisy" database required roughly $10M~CPU-hours$ executed in the NASA Pleiades supercomputer. A larger number of Monte Carlo experiments or simulations would provide more statistical samples (e.g., $\approx$ 100), but would require many more CPU hours -- in the example given $40M CPU-hours$. Such a large number of computational resources is beyond the scope of the current investigation.  

As has been mentioned, the standard procedure would be to consider all (25) \noirma\ to calculate the uncertainties (by determining the standard deviation of the physical parameters determined by the inversions in each experiment). However, due to the limited number of simulations, in some cases, only a few fits out of the 25 fits between the \noirp\ and the resulting inverted profile are "good". In these cases, the standard deviation of these 25 \noirma\ may be very large, since it takes into account a large number of {\it bad} fits. Therefore, we adopt a more empirical approach in which the selection of the \noirma\ considered for calculating the uncertainties is based on the goodness of fit between the \noirp\ and its corresponding inverted profile, i.e., on the value of $\chi^2$:

\begin{equation}\label{eq:noichi2}
\chi^{2} = \frac{1}{\nu}\sum_{i=0}^{q}{(\widetilde{RP}^{syn}_{t_{exp},noi}(\lambda_i) - I(\lambda_i;\   ^{MC}\widetilde{RMA}^{syn}_{t_{exp},noi}))^{2}\frac{w_{i}^2}{\sigma_{i}^2}} 
\end{equation}

 with $\nu$ the number of observables. Note that we are now quantifying the fit between the noisy \noirp\ and its  inverted profile $I$, which is the resulting radiation at $\lambda_i$ from the model \noirma. Thus, for a given \noirp, we have 25 values of $\chi^2$ corresponding to the fits associated between the \noirp\ and the 25 inverted profiles $I$ generated by the 25 \noirma.

\begin{figure*}[t!]
	\centering
	\includegraphics[width=1.\linewidth]{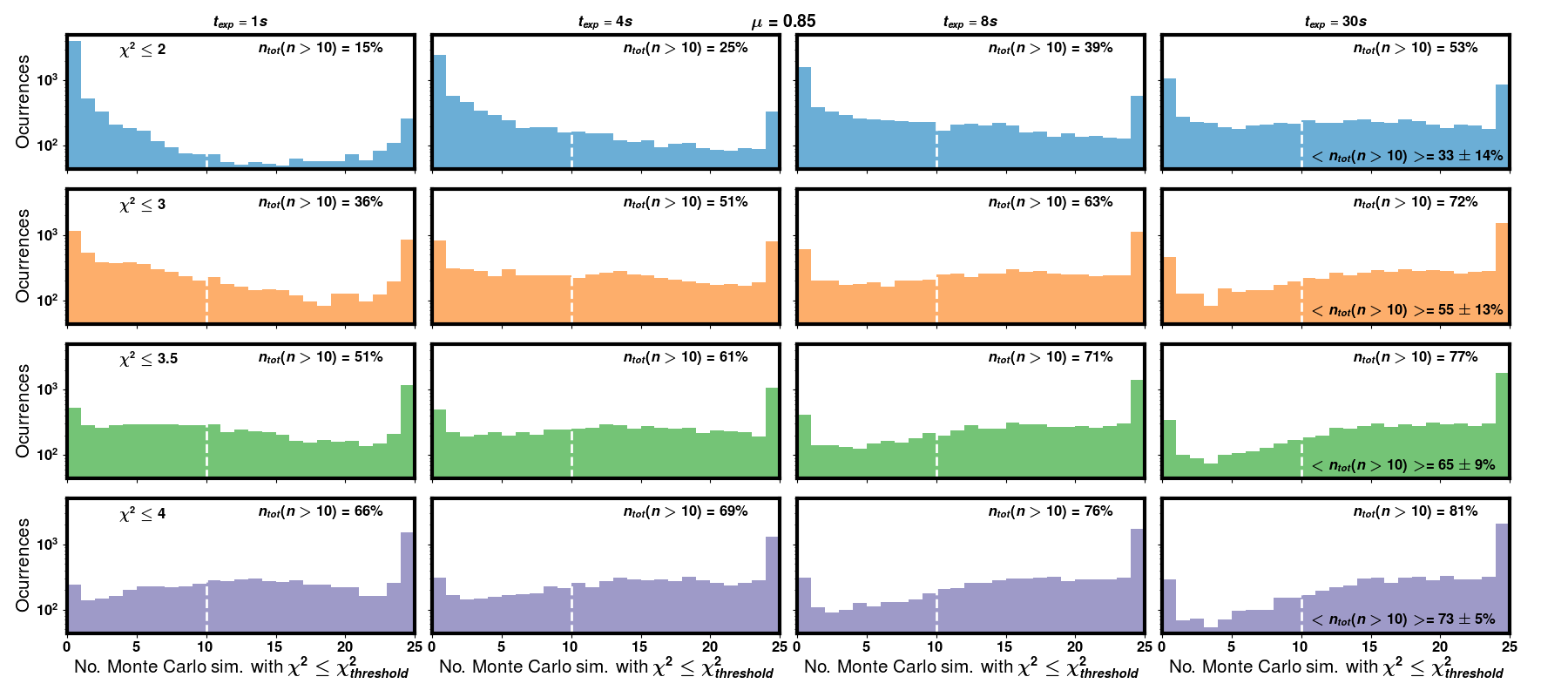}
	\caption{Histograms of number of Monte Carlo simulations (for all \irissq\ entries with $0.8 \le \mu < 0.9$) for which $\chi^2 \le$ \chith, with different values of \chith in each row (indicated in upper left of each panel), and different exposure times in each column. Indicated in the top right of each panel is the fraction of cases for which there are at least 10 Monte Carlo simulations with a goodness of fit better than $\chi \le$ \chith.}
	\label{fig:selchi2}
\end{figure*}

\begin{figure}
	\centering
	\includegraphics[width=1.\linewidth]{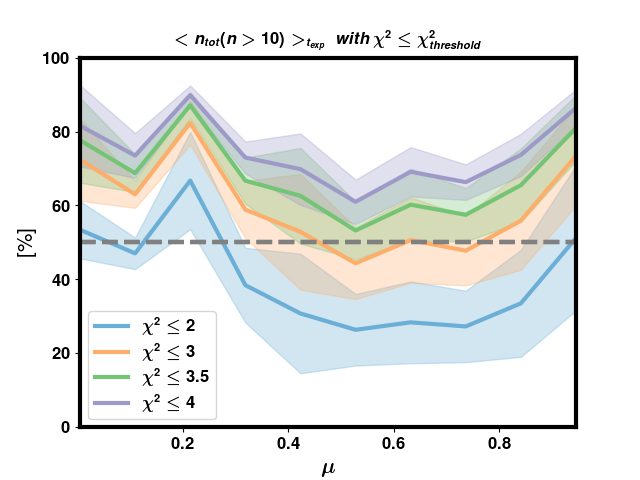}
	\caption{Histogram, as a function of the cosine of the viewing angle ($\mu$), of the average (across all exposure times considered) fraction of \irissq\ database entries for which there are at least 10 Monte Carlo simulations that meet the goodness of fit criterion $\chi \le$ \chith. Colors show different values of \chith, while the ranges shown indicate the standard deviation (across different exposure times) on the average fraction. }
	\label{fig:averchi2all}
\end{figure}

We have selected a criterion that considers a large enough number of experiments to preserve some statistical meaning from the Monte Carlo approach, and that also attempts to minimize the impact of {\it bad} inversions in the calculation of the uncertainties. To enable this, we always consider at least 10 Monte Carlo experiments. If the number of fits with a $\chi^2$ below a given threshold  (\chith)  is less than 10, then the \noirma\ associated with the 10 {\it best} fits are used to calculate the standard deviation of the model. If the number of fits {\it n} with $\chi^2 \le$ \chith\ is larger than 10, then {\it n}  \noirma\ are used to calculated the uncertainties of the asocciated RMA. To justify this empirical approach, we have analyzed the distribution of the number of  inversion fits with a $\chi^2$ below different thresholds. Each row of Figure \ref{fig:selchi2} shows the distribution of the number of fits $n$ with a $\chi^2$ below a threshold (\chith\, indicated in the top left corner in the first column) for each $t_{exp}$ (column)  for the case of $0.8 \le \mu < 0.9$. The threshold values  are \chith$=2,3,3.5$ and $4$.  In each individual panel, the percentage of the total number of $n > 10$ with   with $\chi^2 \le$ \chith\  is indicated in the top right corner, while in the bottom right corner of the last column is indicated the average of these values for all the $t_{exp}$ at a given  $\chi^2 \le$ \chith.  Figure \ref{fig:averchi2all} shows the behavior of the  latter average with respect to $\mu$. In this figure, we can see that for $\chi^2 \le3$, except for $\mu=0.55$ and $\mu=0.75$, the averaged-in-$t_{exp}$ percentage of Monte Carlo experiments (inversions) for a given \synrp\ with at least 10 fits with $\chi^2  \le 3$ is larger than 50\%, and for the values mentioned before the percentages are very close to 50\%. Therefore, we consider \chith$=3$ and $n \ge 10$ to be good criteria to ensure a Monte Carlo simulation with a  well-balanced  number of {\it good} and {\it bad} fits at (almost) any $\mu$ and $t_{exp}$ values. 

%the  Figure \ref{fig:averchi2} shows the average in $t_{exp}$ of the percentages of cases with $n > 10$ fits with $\chi^2 \le$ \chith\ for \chith$=2,3,3.5$ and $4$. We decide that the best option is the one that has a well-balance averaged distribution for all the exposure times, i.e. the one close to $50\%$. 

% TODO: \usepackage{graphicx} required

% TODO: \usepackage{graphicx} required

%As a reference, we also indicate the total number of inversions when the number of fits below the $\chi^2$ set as a threshold is less than 5 and 10. 
%Therefore, the best balanced cases are for $\chi^2 = 3$, which
%Therefore we have decided that the threshold for $\chi^2 = 3.5$ 
%is a good option for capturing the variability of the quality of the inversions at any given exposure time or $\mu$.
 %However, for some cases the number of good fits for a given \noirp may be a small number. In these cases, the standard deviation might lack of statistical meaning. Therefore, we impose a minimum number of 10 fits to calculate the uncertainties. Thus, those \noirp with a number of fits with a $chi^2 \le 3.5$ will have associated a (large) uncertainty from the {\it best} 10 inversions, which will still preserve some statistical meaning.
 In summary, the uncertainty of physical variable $p$ in the RMA is calculated as:
\begin{equation}
\sigma_p = \sigma(^{[N]}\widetilde{RMA}^{syn}_{t_{exp},noi,p})
\end{equation} 
with $[N]$ corresponding to the set formed by the $n$ best fits of the Monte Carlo experiments, which are  determined by:
\begin{equation}
max(n\ with\ \chi^2\le3,\ n=10)\label{eq:condition}
\end{equation} 

For instance, if a \noirp\ has 16 fits with $\chi^2\le3$, then $\sigma_p$ will be calculated considering their 16 associated \noirma, i.e. $^{[16]}\widetilde{RMA}^{syn}_{t_{exp},noi}$. But, if it has only 3 fits with $\chi^2\le3$, then $\sigma_p$ will be calculated considering the corresponding $^{[10]}\widetilde{RMA}^{syn}_{t_{exp},noi}$ to the 10 {\it best} fits, including 7 {\it "bad" fits}, which will result in a larger uncertainty. We believe this approach captures the impact of the uncertainties introduced by the inversion process.  Note that while $MC$ takes values between 1 to 5, $N$ in $[N]$ may take any value from 10 to 25. 

\section{Discussion}\label{sec:discussion}

Let us now discuss the impact of these new calculations on the uncertainties on the thermodynamic parameters from \irissq, and in particular some cases that highlight the difference between the previous and new approach, and the limitations of any uncertainty calculation.

\begin{figure*}[ht!]
	\centering
	%\rotatebox[origin=c]{0}{\includegraphics[width=18cm,height=12cm]{fig_uncertainty_mu85_rp00350_db01619_std25_allatmos_new_big}}%
	\rotatebox[origin=c]{0}{\includegraphics[width=18cm,height=12cm]{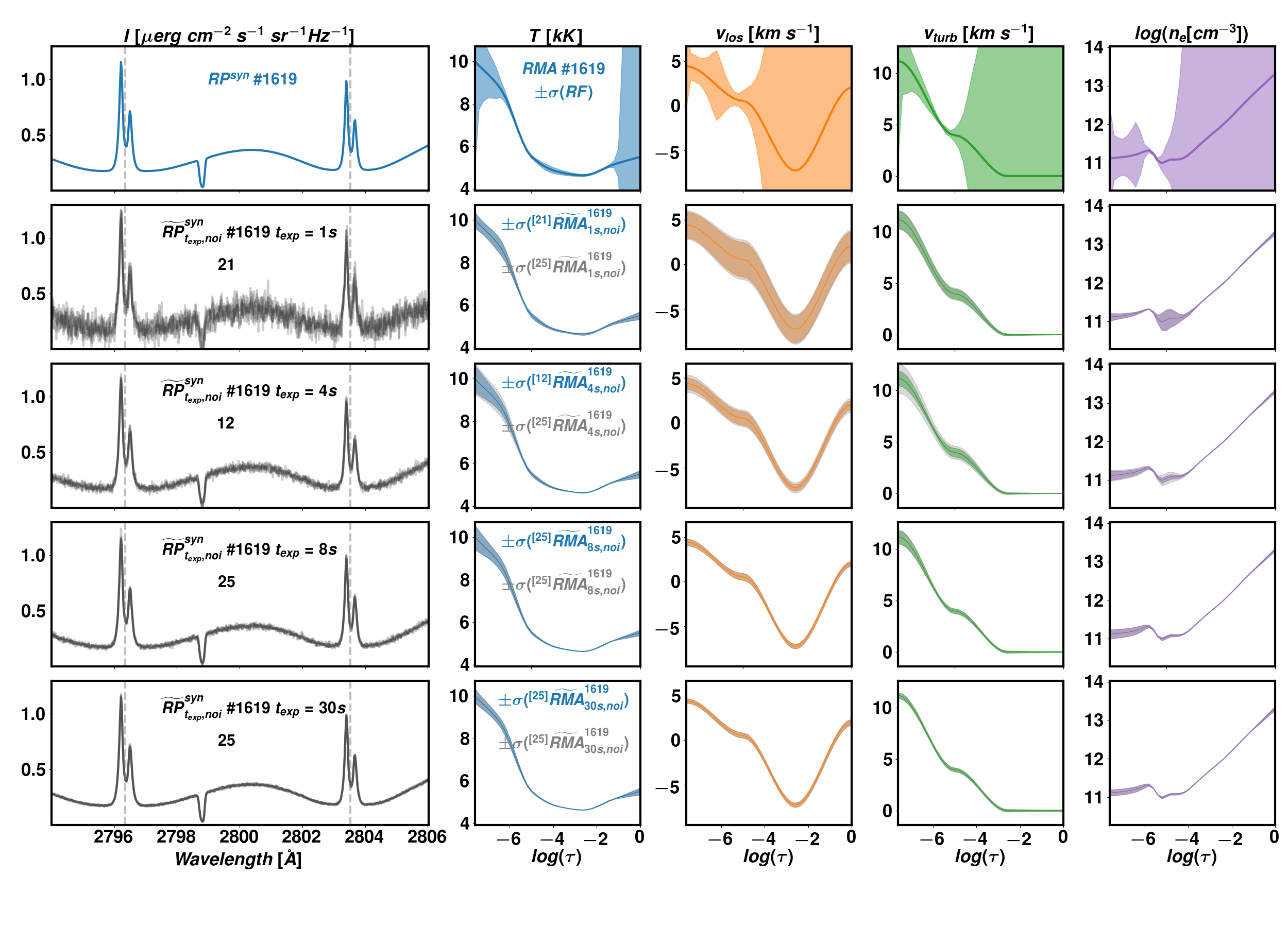}}
	\vspace{-.7cm}
	\caption{The first column shows the \synrp\#1619 in blue, and in black its \noirp for $t_{exp} = 1, 4, 8,$ and $30s$. The corresponding RMA  for $T, v_{los}, v_{turb},$ and $n_e$ and their uncertainties are shown from the second to the fifth column respectively. In the first row, the uncertainty is obtained from the response functions, while the ones from the second to the fifth rows are obtained by using $N$ Monte Carlo experiments as shown in \nnoirma{N}, for  $t_{exp} = 1, 4, 8,$ and $30s$ respectively.}\label{fig:allatmos1}
\end{figure*}

\begin{figure*}[ht!]
	\centering
	%\rotatebox[origin=c]{0}{\includegraphics[width=18cm,height=12cm]{fig_uncertainty_mu85_rp00980_db06395_std25_allatmos_new_big}}%
	\rotatebox[origin=c]{0}{\includegraphics[width=18cm,height=12cm]{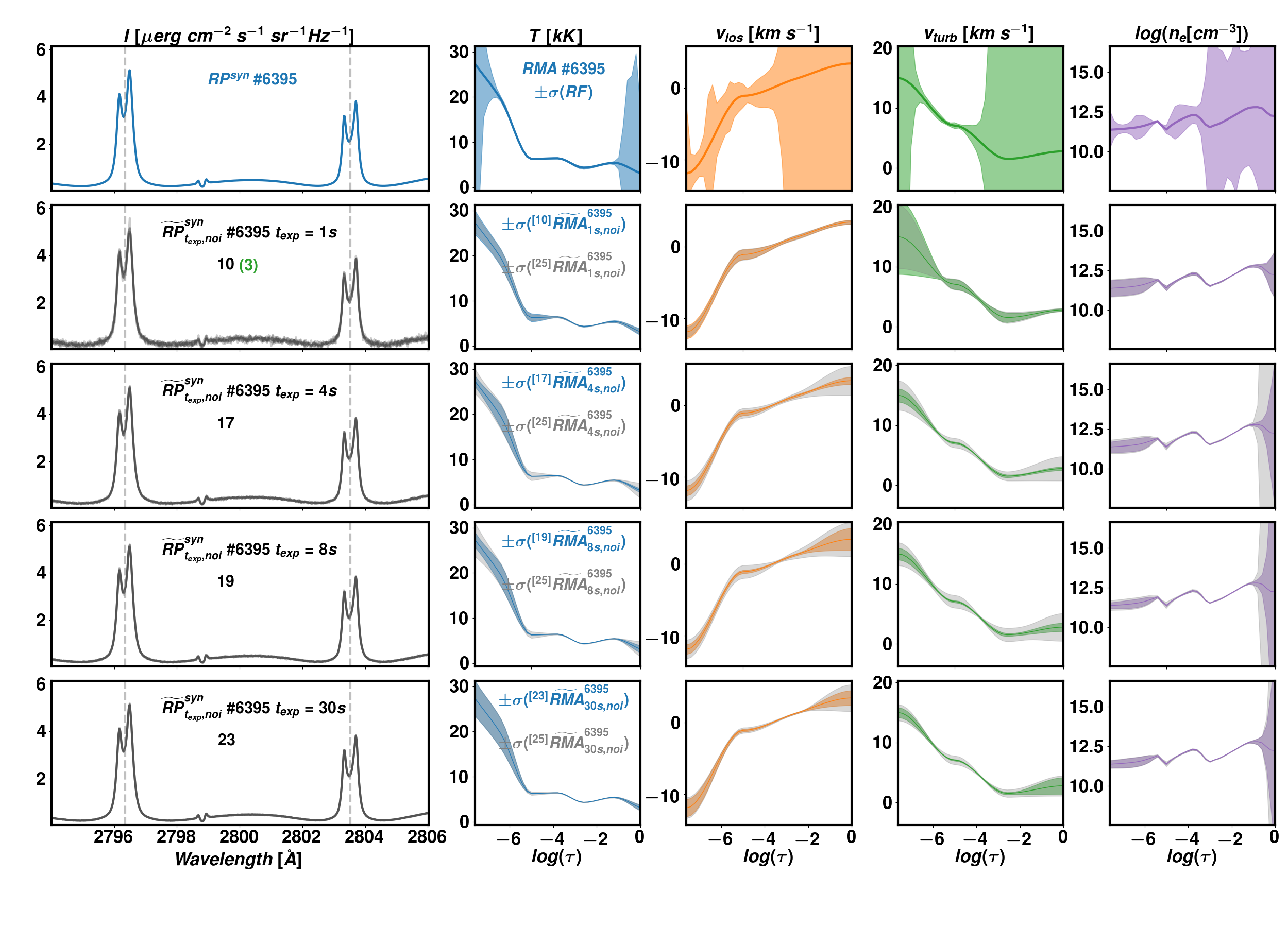}}
	\vspace{-.7cm}
	\caption{Same as Figure \ref{fig:allatmos1} for \synrp\#6395.}\label{fig:allatmos2}
\end{figure*}

The first row of Figures \ref{fig:allatmos1} and \ref{fig:allatmos2} shows the \synrp\ (first column) and the associated RMAs for $T, v_{los}, v_{turb},$ and $n_e$ (in the second, thirth, fourth, and fifth column respectively) with the uncertainties calculated using Eq. 3 in \cite{SainzDalda19}, i.e., using the {\it response functions}. In the following rows, the first column shows \noirp\ for the  \synrp\#1619  with $t_{exp}=1, 4, 8,$ and $30s$, with the 5 noise randomizations over-plotted; from the second to the fifth columns the same RMA thermodynamic variables as in the first row, but now showing two types of uncertainties. In blue, we show the uncertainties calculated using the standard deviation of those \noirma\ associated with the inverted profiles for \noirp\ that satisfy the condition (\ref{eq:condition}). In grey we show the uncertainties derived from all 25 Monte Carlo experiments, i.e., \nnoirma{25}. In each panel of \noirp\ the number of profiles used to calculate the uncertainty is indicated in black, and, as a reference,  the number of profiles that satisfies $\chi^2\le3$ when that number is less than 10 is indicated in green. 

The uncertainties in $T$ are relatively small between $-6 \le log(\tau) \le -3$ for all the $t_{exp}$ in both examples (\#1619 and \#6395). When all the 25 Monte Carlo experiments are considered 
(in grey), we see some differences, with the largest difference at for $-7 \le log(\tau)$ and to a lesser extent around $log(\tau) = -5$.  The former location is the region in the optical depth where neither the \mgii\ nor the \mguv\ are sensitive to the variations in the thermodynamic variables. The latter is where the \mgii\ lines are more sensitive to changes in the atmosphere. Therefore, we should expect some uncertainty in the atmosphere for inversion cases in which the \noirp\ are not well fitted, and also where the \mgii\ lines  are actually sensitive to variations in the thermodynamic parameters. For  $-3 \le log(\tau)$, the uncertainties are usually larger, which makes sense since the IRIS \mgii\ profiles barely encode photospheric information, i.e., these lines are not sensitive to variations in the thermodynamics at this optical depth range. 

It is important to distinguish how the uncertainties are calculated in the considered methods. In the method using the RFs, a small variation in the atmospheric parameter  is introduced at given optical depth, then the {\it response function} is obtained as the difference between the synthetic profile from the atmosphere with the slightly modified parameter with respect to the profile corresponding to the unperturbed model atmosphere (i.e., without variation of any physical parameter). This process is repeated for all the optical depths considered in the model atmosphere. Let us now consider how uncertainties are determined in our new method. First, we note that during the inversion of the profiles only some optical depths (nodes\footnote{The cycles and nodes used in this study are the same as the ones used in \citealt{SainzDalda19}: the first
cycle considers four nodes in temperature, and three nodes
both $v_{turb}$ and $v_{los}$.
The second cycle uses seven nodes in temperature, and four nodes
both in $v_{turb}$ and $v_{los}$.}) are considered. In the Monte Carlo approach, five full inversions for the five \noirp\ are executed to evaluate the reproducibility of the results, using the standard deviation of the resulting models as the uncertainties for the original model. In the first case (using RFs), the synthetic profiles come from a model atmosphere evaluated in all the optical depths with a small variation, while this is not the case in our new calculations: the variation of the model atmosphere during the inversion only occurs in the nodes. In some cases, the inversion code may find a good fit generating some variations in some nodes, and none (or negligible) in other nodes because the code is able to fit the input profile without variation in these nodes. This is why in some cases the uncertainties at  $-2 \le log(\tau)$ are very small. This effect can be seen for $v_{los}$ for \noirma\#6395 with $t_{exp}=1s$ in comparison with the other $t_{exp}$. For the latter, the \mguv lines are more well defined (less noisy) and the inversion code may be trying to introduce a variation in the nodes at $-3  \le log(\tau)$. This effect is also noted in the $v_{turb}$,  and in the $n_e$ at  $-1  \le log(\tau)$.  The conclusion then is that when assessing the uncertainties, we have to be aware of the optical depths where the observed lines are mostly sensitive to different parameters. These regions are slightly different for different solar features (e.g. umbra, penumbra or plage), and different between the physical parameters (e.g., see Fig. 2 in \citealt{delaCruzRodriguez16}). 

\begin{figure*}
	\centering
	\includegraphics[width=\textwidth]{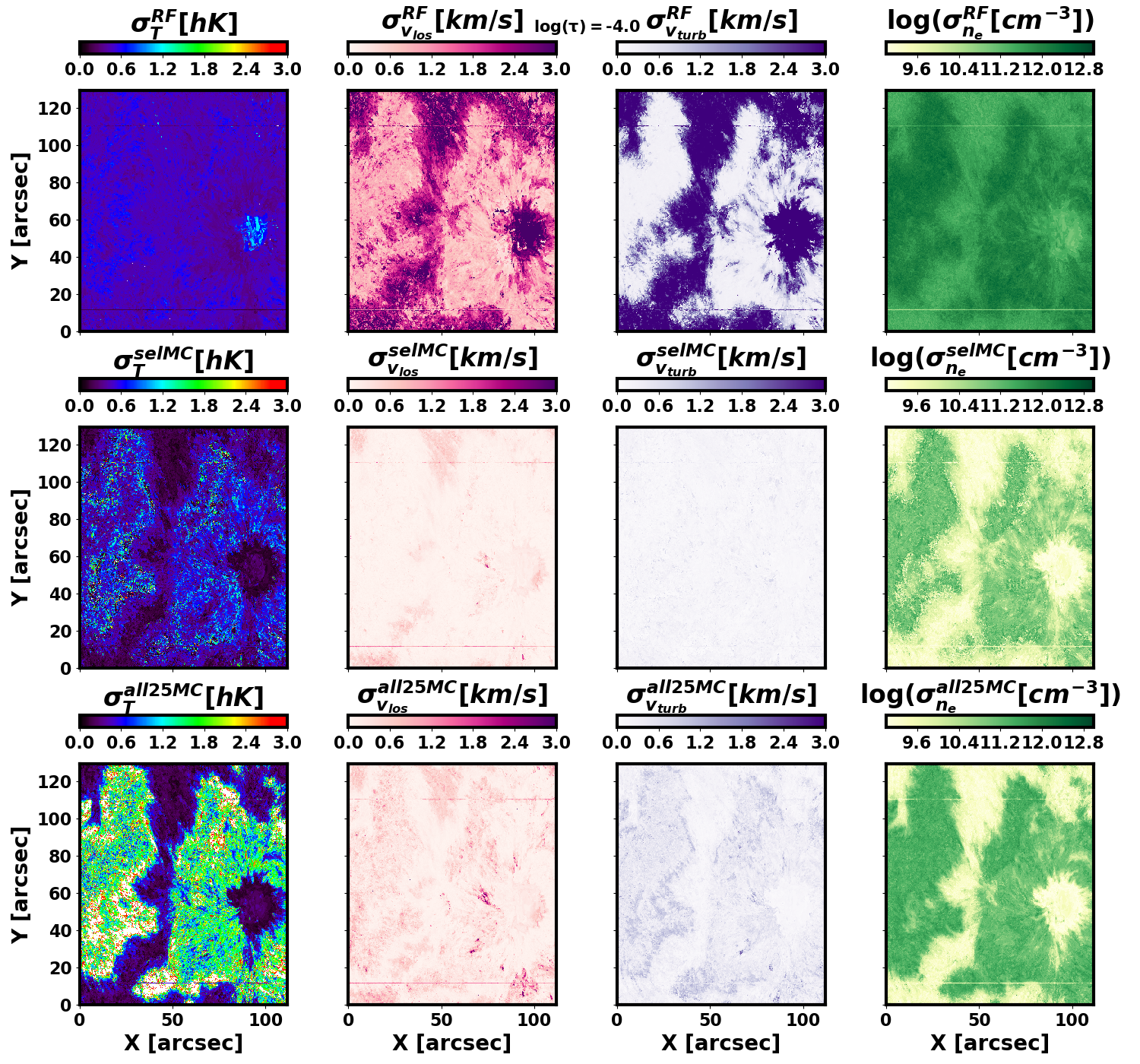}
	\caption{The uncertainties for the $T, v_{los}, v_{turb},$ and $n_{e}$ at $\log(\tau) = -4$ when considering the RFs (top row), the {\it selective} Monte Carlo approach (middle row), or all the 25 Monte Carlo experiments (bottom row).}
	\label{fig:figuncertltau-4}
\end{figure*}

\begin{figure*}
	\centering
	\includegraphics[width=1\linewidth]{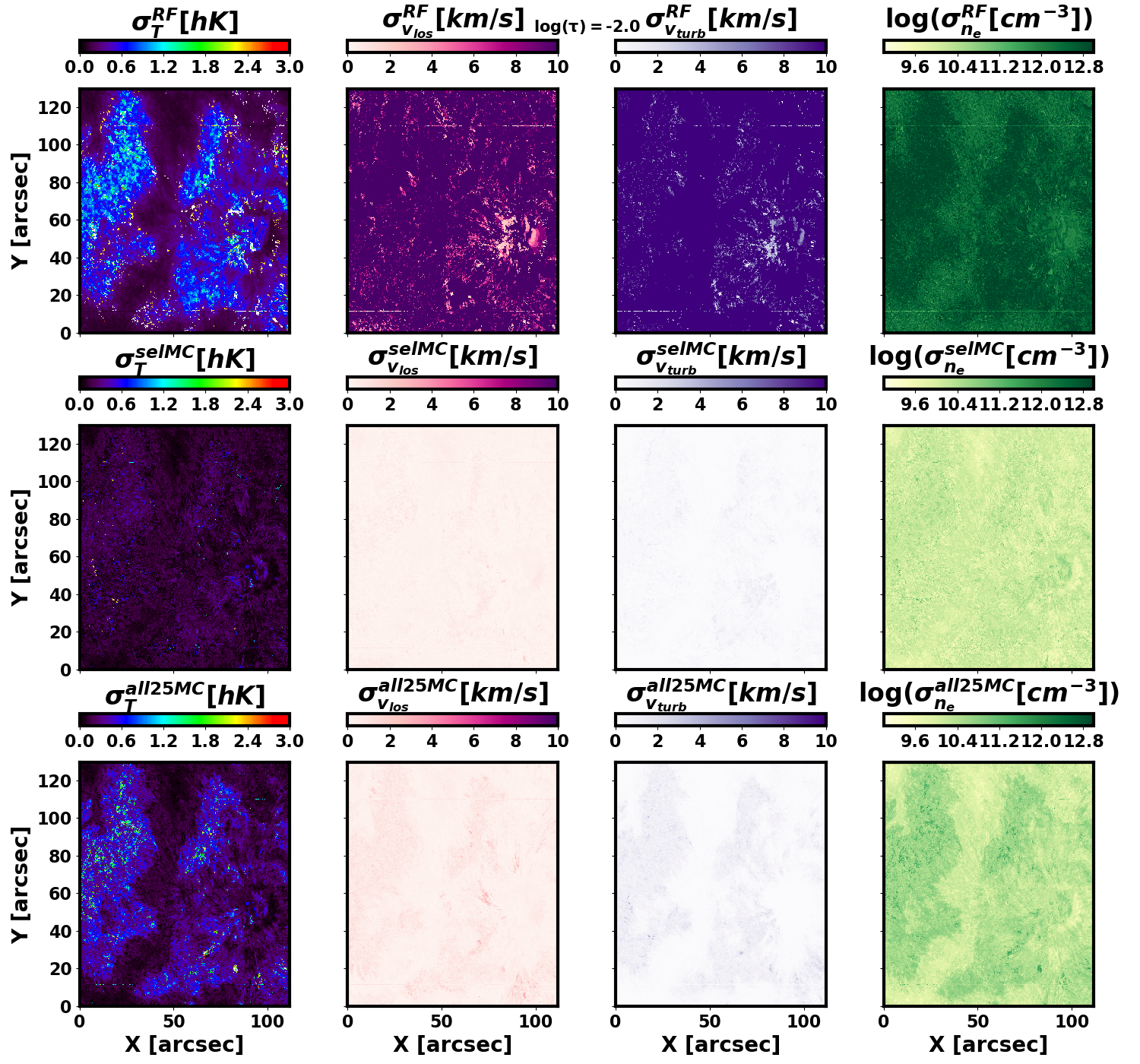}
	\caption{Same as Figure \ref{fig:figuncertltau-4} for $\log(\tau) = -2$.}
	\label{fig:figuncertltau-1}
\end{figure*}

The first row of Figure \ref{fig:figuncertltau-4} shows maps of the uncertainty calculated using the RFs ($\sigma^{RF}$) of $T, v_{los}, v_{turb},$ and $n_{e}$ at $\log(\tau) = -4$. The second and the third rows show respectively the uncertainties calculated using the {\it selective} Monte Carlo experiments ($\sigma^{selMC}$) , i.e. \nnoirma{N},  and all 25 Monte Carlo experiments ($\sigma^{all25MC}$), i.e. \nnoirma{25}. At this optical depth, in the plage and the umbra and extended penumbra or canopy the $\sigma^{RF}_T <  \sigma^{selMC}_T <<  \sigma^{all25MC}_T$, while for the $v_{los}$, $v_{turb}$ , and $n_e$  the $\sigma^{RF} > \sigma^{all25MC} >> \sigma^{selMC}$. This situation is however different at $\log(\tau) = -2$ (see Figure \ref{fig:figuncertltau-1}), where  $\sigma^{RF}_T > \sigma^{all25MC}_T >>  \sigma^{selMC}_T$, and for the  $v_{los}$, $v_{turb}$ , and $n_e$  the $\sigma^{RF} >> \sigma^{all25MC} >> \sigma^{selMC}$. These two figures illustrate what we mentioned above. When we calculate uncertainties from the response functions (as in \citealt{SainzDalda19}), the uncertainties may be too low for those optical depths where the lines are sensitive to changes in the thermodynamics (large RFs), while they may be unrealistically high for those optical depths where the line is barely sensitive to changes in the thermodynamics (small RFs). We find that, for the Monte Carlo approach, the variation with optical depth of the uncertainties is more moderated, and typically smaller for the selective criterion than when considering all 25 Monte Carlo experiments. 
%However, that is not the case either for  \synrp\#1619 or - more notoriously - for \synrp \# 6395 (

%\begin{sidewaysfigure*}[h]
%	\centering
%	\includegraphics[width=\textwidth]{example-image}
%	\caption{Intensity measure correlation}
%	\label{fig:IMs matrix correlation}
%\end{sidewaysfigure*}

\section{Conclusions}\label{sec:conclusions}

In this paper, we present and discuss a novel methodology and the results of applying a selective Monte Carlo approach to determine the uncertainties associated with the {\it Representative Model Atmosphere} (RMA) in the \irissq\ data base. These new uncertainties represent more realistic values than the previously publicly released uncertainties \citep{SainzDalda19} which were based on response functions. This is because the uncertainties in our new approach have been calculated from the synthetic representative profiles \synrp\ considering the different sources of uncertainty in the whole process, i.e.: different exposure times, different noise randomization, and different inversion initializations. We define the uncertainty of a  physical parameter  associated with the pair \synrp-RMA as the standard deviation of this parameter in the set of depth-stratified output models (from the Monte Carlo experiments) that satisfy the {\it ad hoc} selection criterion shown in Eq. (\ref{eq:condition}). The latter expression is used to minimize the impact of the output models based on inversions that produce a bad fit (with the  noisy synthetic profile associated with \synrp). In general, at the optical depths where the \mgii\ and \mguv\ lines are sensitive to variations in a thermodynamic parameter, the difference between considering all 25 Monte Carlo experiments instead of the number that satisfies  expression (\ref{eq:condition}) is very small. The difference is larger for optical depths where the lines are not sensitive to thermodynamic changes. %, being, in general, this difference larger in those optical depths where these lines are not sensitive to changes in the thermodynamic.

The new uncertainties will be available to the public in the \irissq data base, both for IDL and Python. The uncertainty calculated from the 25 Monte Carlo experiments will also be provided as an extra field  in the new version of the \irissq data base. Therefore,  the new data base will have the following elements:
\begin{itemize}
	\item \synrp: 472 wavelength positions, from 2794 to 2806\AA, with a spectral sampling of $\approx 0.025m$\AA
	\item RMA: depth-stratified $T, v_{los}, v_{turb},$ and $n_e$, sampled at 39 optical depths (i.e., "heights" in the atmosphere) with $\delta log(\tau)=0.2$
	\item $\sigma_{sel}$: depth-stratified $\sigma_{T}, \sigma_{v_{los}}, \sigma_{v_{turb}},$  and $\sigma_{n_e}$, sampled at 39 optical depths with $\Delta(log(\tau))=0.2$, obtained from the selected Monte Carlo experiments ({\it selective mode}). These values are given for $t_{exp} = 1, 4, 8,$ and $30s$. Therefore, $4 \times \sigma_{sel}$ values are in the database. 
   \item $\sigma_{all} $: depth-stratified $\sigma_{T}, \sigma_{v_{los}}, \sigma_{v_{turb}},$ and $\sigma_{n_e}$, sampled at 39 optical depths with $\Delta(log(\tau))=0.2$, obtained from the 25 Monte Carlo experiments ({\it all-in mode}). These values are given for $t_{exp} = 1, 4, 8,$ and $30s$. Therefore, $4 \times \sigma_{all}$ values are in the database. 
   \item $\mu$: from 0 to 1,  starting from $\mu=0.05$ at steps of 0.10, as indicated in Table \ref{table:mu}.
\end{itemize}

%The \irissq database will be available to the public as described above. 
The different \irissq inversion tools that allow users to interface with this database will use these database elements for internal calculations. The inversion of an IRIS \mgii\ data set will only return the closest \synrp\ to the observed profiles, the corresponding RMAs, and the uncertainties taking into account the $\mu$ and the $t_{exp}$ of the observation and the uncertainty mode ({\it selective} or {\it all-in}) chosen by the user. 

We believe that the empirical methodology we have developed for \irissq will be useful for understanding the uncertainties associated with other or similar inversion approaches.
%\begin{figure}
%	\centering
%		\includegraphics[angle=90, scale=.5]{figures/fig_uncertainty_mu85_rp00980_db06395_std25_new}
%%      \includegraphics[angle=90, scale=0.5]{figures/fig_uncertainty_mu85_rp00980_db06395_std25_new}
%	\caption{}
%	\label{fig:figinvnoisymu85rp00350db01619std25new}
%\end{figure}
%\begin{figure}
%	\centering
%%	\includegraphics[angle=90, scale=0.5]{figures/fig_uncertainty_mu85_rp00980_db06395_std25_new}
%	\includegraphics[angle=90, scale=0.5]{figures/fig_uncertainty_mu85_rp00980_db06395_std25_new}
%	\caption{}
%	\label{fig:figinvnoisymu85rp00350db01619std25new}
%\end{figure}

\acknowledgements
IRIS is a NASA small explorer mission developed and operated by LMSAL with mission operations executed at NASA Ames Research center and major contributions to downlink communications funded by ESA and the Norwegian Space Agency. This work was supported by NASA contract NNG09FA40C (IRIS). Resources supporting this work were provided by the NASA High-End Computing (HEC) Program through the NASA Advanced Supercomputing (NAS) Division at Ames Research Center. The inversions were run on the Pleiades cluster through the computing project s1061 from the NASA HEC program. The authors are grateful to Andrés Asensio Ramos and Jaime de la Cruz Rodríguez for insightful discussions, and to Marc DeRosa for his improvements in the text.
%\facility{facility ID}
\software{\irissq, see \url{https://iris.lmsal.com/iris2/}}

\appendix
\begin{minipage}{\textwidth}
\section{Limitations of inversion approach}

This appendix describes in more detail some limitations of the inversion approach that has been used for the IRIS$^2$ database.

As we mentioned above, Figure \ref{fig:invnoisy} shows the \synrp (top row), its 5 associated \noirp\  (black)  and its 25 associated inverted profiles (violet and orange for good and bad fits respectively)  for two cases of the \irissq database. 

For the \synrp \#1619 (top panel), we can see an interesting behavior: the number of good fits for $t_{exp}=1s$ is almost as large as for $t_{exp}=8$ and $30s$, and definitely larger than for $t_{exp}=4s$. At first glance, one would perhaps expect that for longer $t_{exp}$ finding a good inverted profile close to the \noirp\, should be more difficult than for a profile with shorter $t_{exp}$. However, the latter is noisier than the former, and thus it has a larger variation in its values, making it easier to find a fit that is good enough to end the iterative process of the inversion and for the code to declare a "good fit". This is easily visible in the second row of the top panel: larger noise in the \noirp\ allows the  inverted profile to fit more easily to the \noirp. That means, the difference between the \noirp\ and the candidate to the final inverted profile is less than the noise. The larger the noise, the easier the expression (\ref{eq:condition}) can be satisfied. However, during the inversion process, the code may find a local minimum in the search for the best fit, and therefore it may not able to find a better solution, and eventually reach the number of maximum iterations allowed. On the other hand, it can also occur that the code actually finds the best fit in all the cases despite smaller noise, as seems to happen for $t_{exp}=8$ and $30s$. This is even more evident for the \noirp\ \#6395. %Nevertheless, the situation here is slightly different. 

We now describe another peculiarity related to the inversions.
During the inversion, the code tries to minimize the $\chi^2$,  which is basically the average of the ratio of the weighted difference of the \noirp\ and the fit from the inversion with respect to the noise.
As we have already mentioned, because of computational constraints the inversion code only accepts a single noise value for all the profiles considered in the inversion. That means, the noise is the same at any wavelength. And more importantly, it is the same for a profile where the ratio between the line (peaks and the core) and the photospheric bump ($r_{l2b} = I_{line}/I_{bump}$) is large (e.g., a location with strong chromospheric heating such as the \synrp\ \#6385) as for a profile with a small $r_{l2b}$ (e.g., a quiet Sun location such as  the \synrp\ \#1619).  Whether the core of the spectral lines (as opposed to the wings or continuum) has a large impact on the $\chi^2$ value depends on the value for $\r_{l2b}$, the noise value, and the number of wavelengths sampled within and outside of the wavelength range covered by the spectral lines\footnote{Both in the inversions used to build \irissq\ and the ones used in this current work, the weights of the lines, photospheric bump, and wings are taken to be the same.}.
%In those profiles with a large $r_{l2b}$, the difference between the \synrp\ and the \noirp\ in the lines (peaks and core) may have a large impact in the $\chi^2$ than in the bump, even when the number of spectral positions (in white circles in the \synrp\ in Figure  \ref{fig:invnoisy}) is relatively small in the lines. 

Thus, if the noise is large (e.g. for $t_{exp}$ equal to 1 or 4 $s$) and the $r_{l2b}$ is small, the contribution of the lines and the bump to the $\chi^2$ is very similar, since the difference between the \synrp\ and the \noirp\ in the line and the bump are similar. For that reason the \synrp \#1619 has a large number "good" fits for short $t_{exp}$: there are large number of \synrp\ that on average fit the \noirp\ within the (large) noise, even when the core of the lines is not well fit, since the contribution of the small number of sample wavelengths in the line to the $\chi^2$ is small. However, if the noise is large but the $r_{l2b}$ is large, since the values in the line are much larger than in the bump, they will have a significant impact in the $\chi^2$. Therefore, $\chi^2$ will {\it more easily} consider as "bad" fits those profiles that have a poor fit in the line (usually in the core). This is the case for \synrp\ \#6385 for $t_{exp}$ is 1 or 4 $s$. If the noise is small, all the points both in the line and the bump have to fit more strictly, since the difference between the \synrp\ and the \noirp\ should be comparable to the small noise. In this case, since the noise is small, the inversion will look for solutions that strictly fit all the sampled wavelengths of the \noirp\: both the line and the bump have a similar impact in the $\chi^2$. This happens in the \synrp\ \#1619 and \#6385 when $t_{exp}$ is 8 or 30 $s$.

In summary, we can see that $\chi^2$ is not necessarily always the best metric (or loss function)
to quantify the quality of the fit of \noirp\ in the inversions. This is due to the high dimensionality of the profiles (a large number of sampled spectral positions) and the computational constraints that impose the same  weight and noise per spectral sample and per \synrp\ and per data set in \irissq\ in this study. 
\end{minipage}

%\bibliographystyle{yahapj}
%\bibliography{allbib,others}

\end{document}